%Paper: hep-th/9501072
%From: wisskirc@avzw01.physik.uni-bonn.de (Andreas Wisskirchen)
%Date: Wed, 18 Jan 95 09:27:38 +0100

%%% Macros etc.
\magnification=\magstep1
\overfullrule=0pt

\def\w{$\cal W$}
\font\klein=cmr8 scaled\magstep0

\def\q#1{\lbrack #1 \rbrack}
\def\smno{\smallskip\noindent}
\def\meno{\medskip\noindent}
\def\bigno{\bigskip\noindent}
\def\pano{\par\noindent}

\def\o#1{\overline{#1}}

\def\pt{\partial}

\def\lb{\lbrack}
\def\rb{\rbrack}

\def\ts{\textstyle}
\def\cl{\centerline}

\def\id{1\hskip-2.5pt{\rm l}}
\def\BNT{\,\hbox{\hbox to -2.2pt{\vrule height 6.5pt width .2pt\hss}\rm N}}
\def\BZT{{\rm Z{\hbox to 3pt{\hss\rm Z}}}}
\def\BZS{{\hbox{\sevenrm Z{\hbox to 2.3pt{\hss\sevenrm Z}}}}}
\def\BZSS{{\hbox{\fiverm Z{\hbox to 1.8pt{\hss\fiverm Z}}}}}
\def\BZ{{\mathchoice{\BZT}{\BZT}{\BZS}{\BZSS}}}
\def\BQT{\,\hbox{\hbox to -2.8pt{\vrule height 6.5pt width .2pt\hss}\rm Q}}
\def\BQS{\,\hbox{\hbox to -2.1pt{\vrule height 4.5pt width .2pt\hss}$
    \scriptstyle\rm Q$}}
\def\BQSS{\,\hbox{\hbox to -1.8pt{\vrule height 3pt width
    .2pt\hss}$\scriptscriptstyle \rm Q$}}

\def\BCT{\,\hbox{\hbox to -3pt{\vrule height 6.5pt width .2pt\hss}\rm C}}
\def\BCS{\,\hbox{\hbox to -2.2pt{\vrule height 4.5pt width .2pt\hss}$
    \scriptstyle\rm C$}}
\def\BCSS{\,\hbox{\hbox to -2pt{\vrule height 3.3pt width
    .2pt\hss}$\scriptscriptstyle \rm C$}}

\def\section#1{\leftline{\bf #1}\vskip-7pt\line{\hrulefill}}
\def\bibitem#1{\parindent=8mm\item{\hbox to 6 mm{$\q{#1}$\hfill}}}
%%% Authors
\def\abel{1}
\def\ant{2}
\def\banks{3}
\def\self{4}
\def\cande{5}
\def\candz{6}
\def\candd{7}
\def\egu{8}
\def\fuchs{9}
\def\gepe{10}
\def\gepz{11}
\def\gre{12}
\def\gros{13}
\def\kaw{14}
\def\kreuz{15}
\def\lere{16}
\def\lerz{17}
\def\oda{18}
\def\sche{19}
\def\schz{20}
\def\schd{21}
\def\rolfe{22}
\def\rolfz{23}
\def\rolfd{24}
\def\terry{25}
%%% Fonts for title page

\font\Large=cmr12 scaled \magstep3

\footline{\hss\tenrm\folio\hss}
\rm
%%% Title Page
\nopagenumbers
\pano
{\rightline {\vbox{\hbox{hep-th/9501072}
                   \hbox{BONN-TH-95-03}
                   \hbox{IFP-503-UNC}
                   \hbox{January 1995} }}}
\bigno\bigno
\centerline{\Large Generalized String Functions Of N=1}
\vskip 10pt
\centerline{\Large Space-Time Supersymmetric String Vacua}
\vskip 1.0cm
\centerline{R.\ Blumenhagen${}^1$ and A.\ Wi{\ss}kirchen${}^2$}
\vskip 1.0cm
\centerline{${}^1$ \it Institute of Field Physics, Department of Physics
and Astronomy}
\centerline{\it University of North Carolina, Chapel Hill NC 27599-3255, USA}
\vskip 0.1cm
\centerline{${}^2$ \it Physikalisches Institut der Universit\"at Bonn,
Nu{\ss}allee 12, 53115 Bonn, Germany}
\vskip 1.0cm
\centerline{\bf Abstract}
\meno
We present a dual formulation of the construction of $N=2$ nonlinear
$\sigma$-model type conformal field theories with $c=9$ which are mainly
used as internal sectors of Calabi-Yau heterotic string compactifications.
The supercurrents $G^\pm(z)$ and the higher components of the spectral flow
superfields $X^\pm(z)+\theta^\pm Y^\pm(z)$ turn out in each case to be
tensor products of two simple currents of a $c=1$ and a non-supersymmetric
parafermionic $c=8$ CFT. The characters of the latter model can be
regarded as string functions of the coset ${\sigma\over U(1)}$.
In particular, for the $(1)^9$ Gepner model we discuss these
string functions in detail, realizing this model  contains a broken
$E_8$ gauge symmetry of which only the abelian subalgebra remains a
symmetry of the spectrum. As an example, we construct a simple model
leading to a string theory with a massless spectrum of 36 $E_6$ generations
and an extended gauge symmetry $E_6\times SU(3)^4$.
\footnote{}
{\pano
${}^1$ e-mail: blumenha@physics.unc.edu
\pano
${}^2$ e-mail: wisskirc@avzw01.physik.uni-bonn.de}
\vfill
\eject
%%% text
\footline{\hss\tenrm\folio\hss}
\pageno=1
\section{1.\ Introduction}
\meno
Since the realization of the importance of Calabi-Yau manifolds and $N=2$
world-sheet conformal field theories (CFT) for the construction of
phenomenologically interesting four-dimensional $N=1$ space-time
supersymmetric heterotic string models [\gros], a large set of consistent
models has been investigated
[\ant,\banks,\cande,\fuchs,\gepe,\kaw,\lere,\lerz],
including many three generation cases [\abel,\candz,\gepz,\schd,\rolfe].
In 1988 D.\ Gepner [\gepe] suggested a quite interesting construction of
string vacua which could in principle be performed for every $N=2$ CFT with
central charge $c=9$. The main ingredient is an automorphism of the $N=2$
Virasoro algebra, the so-called spectral flow, which naturally provides one
with the space-time supersymmetry generator and a GSO projection on the
world-sheet. The authors of [\egu] mainly focused their attention on CFTs
describing $N=2$ nonlinear $\sigma$-models on Calabi-Yau spaces. These
$\sigma$-models could be used as the internal sector of heterotic string
models leading to a minimal gauge group $E_6$. The characters of a
$\sigma$-model CFT are given by flow invariant orbits of integer $U(1)$
charge of some arbitrary rational $N=2$ CFT. Their actual idea to use
directly the extension of the $N=2$ Virasoro algebra by the local spectral
flow operators $X^{\pm 3}(z)+\theta^\pm Y^{\pm 2}(z)$ of superconformal
dimension $(H,Q)=(3/2,\pm 3)$, a nonlinear $\cal W$-algebra, however, failed
because of the non-rationality of this algebra [\oda]. In this letter we will
show that using only a rational subalgebra of the entire spectral flow
$\cal W$-algebra, one arrives at a dual formulation of $\sigma$-model CFTs,
in which the roles of the spectral flow $X^{\pm 3}(z)$ and the supercurrent
$G^\pm(z)$ are interchanged. The $c=9$ theory is split into two parts of
central charge $c=1$ and $c=8$, respectively. The former one contains the
$U(1)$ current and the original spectral flow operators $X^{\pm 3}(z)$,
whereas the latter part is non-supersymmetric. The characters of this $c=8$
CFT can be regarded as generalized string functions of the $\sigma$-model
with respect to the $U(1)$ current. After describing this in more detail in
sect.\ 2, we calculate the string functions for the $(1)^9$ Gepner model,
revealing a hidden $E_8$ structure which can be used to reconstruct this
model in the dual approach using only $E_8$ $\Theta$-functions at level
$k=3$. Finally, in sect.\ 4 we constructively apply the dual formalism to a
model where the $c=8$ theory is given by four copies of the $SU(3)_1$
Kac-Moody algebra. The model turns out to have 36 $E_6$ generations and an
enlarged gauge group $E_6\times SU(3)^4$. Since our construction is quite
general, all methods used for building $N=1$ space-time supersymmetric
string models should fit into this scheme, e.g.\ the latter fairly simple
model appears in a rather complicated covariant lattice construction [\lerz].
\bigno
\section{2.\ Dual  construction of $N=2$ $\sigma$-model partition functions}
\meno
In this section we describe a general formalism to construct $N=2$
supersymmetric $\sigma$-models of Calabi-Yau manifolds.
\footnote{${}^{\dag}$}{Recently, it has been pointed out that in order to
allow also mirror manifolds of rigid models with $n_{\o{27}}=0$ one has to
consider more general classes of manifolds, so-called special Fano varieties
[\candd,\rolfz,\rolfd].} It is partially a review of the work of Eguchi,
Ooguri, Taormina, Yang [\egu], however, with more emphasis on rational
modular properties of the characters involved. First, we recall the important
role played by the spectral flow algebra, which leads to a quite general form
of that part of the $\sigma$-model characters which depends explicitly on
the $U(1)$ charge $Q$. Secondly, knowing the $\sigma$-model partition
function in the $NS$ sector, we show how to construct a consistent $N=1$
space-time supersymmetric string model from it.
\pano
Suppose, there is a given rational $N=2$ super CFT with central charge
$c=9$ and characters $\chi^{NS}_i(\tau,\theta)$ which form a
finite-dimensional representation of the subgroup $M'$ of the modular group
generated by $\{ T^2,S\}$. The continuous automorphism of the $N=2$ Virasoro
algebra
$$\eqalignno{ &L'_n=L_n+\eta J_n +{c\over 6}\eta^2\delta_{n,0}
            \quad\quad   (G^\pm)'_n=G^\pm_{n\pm\eta}
            \quad\quad J'_n=J_n+{c\over 3}\eta\delta_{n,0}&(2.1)\cr}$$
is called the spectral flow. For $\eta=1$ its action on the representations
of the $N=2$ CFT arranges them into a finite number of spectral flow
invariant orbits. It can be shown [\gepe] that for $c=3d\,,d\in\BNT$ all
orbits with integer $U(1)$ charge form a subrepresentation of $M'$ and
restriction of one modular invariant partition function onto these orbits
yields another one. It has been pointed out in [\egu] that these new
invariants are the partition functions of $N=2$ nonlinear $\sigma$-models on
Calabi-Yau manifolds. Implementing these as the internal part of a heterotic
string model yields vacua with phenomenologically desired $N=1$ space-time
supersymmetry. We remark that the construction of new modular invariants
outlined above can also be considered as modding out the spectral flow
simple current $Y^{+2}(z)$ with  superconformal weights $(H,Q)=(2,2)$ using
the Schellekens-Yankielowicz technique [\sche,\schz]. In this case, the
monodromy charge is equal to the $U(1)$ charge. Furthermore, the extension
of the $N=2$ Virasoro algebra by the local, holomorphic and (anti)chiral
super fields $X^{\pm 3}_{3/2}(z)+\theta^{\pm} Y^{\pm 2}_2(z)$ yields a
nonlinear \w-algebra. Thus, one can also consider the spectral flow
invariant orbits as  general reducible representations of this
${\cal{SW}}(1,(3/2)^{\pm 3})$ algebra. The main obstacle for using this
extended algebra for the direct construction of new models is its
non-rationality. Although only a finite number of $U(1)$ charges $Q$ are
allowed, there are no restrictions on the values of the conformal dimensions
$H$. \footnote{$\ddag$}{Note that the involved states need not to be
(anti)chiral.} The idea is to use not the complete spectral flow algebra,
but only a rational subalgebra. The ${\cal{SW}}(1,(3/2)^{\pm 3})$ algebra
contains two $N=2$ Virasoro algebras, the obvious one with central charge
$c=9$ and another one with $c=1$, the generators of which are
$$ {\ts \left\{ {1\over 3}J, {1\over 6}J^2,
{1\over 3}X^+, {1\over 3}X^- \right\}.}\eqno(2.2) $$
Since this algebra admits only three representations in the $NS$ sector with
weights $(H,Q)\in\{ (0,0),(1/6,\pm 1/3)\}$, the flow invariant orbits
$\chi^{\sigma,NS}_i$ can be expanded into three terms:
$$\eqalignno{ \chi^{\sigma,NS}_i(\tau,\theta)=&
 \chi^{N=2,c=1}_{0,0}(\tau,3\theta) \, S_i(\tau)+
     \chi^{N=2,c=1}_{1/6,1/3}(\tau,3\theta) \, T_i(\tau)+
     \chi^{N=2,c=1}_{1/6,-1/3}(\tau,3\theta) \, U_i(\tau) &\cr & &(2.3)\cr
 =&{\ts {\Theta_{0,6}(\tau,\theta)+\Theta_{6,6}(\tau,\theta)\over
                              \eta(\tau)}}\, S_i(\tau)+
          {\ts {\Theta_{-4,6}(\tau,\theta)+\Theta_{2,6}(\tau,\theta)\over
                              \eta(\tau)}}\, T_i(\tau)+
          {\ts {\Theta_{-2,6}(\tau,\theta)+\Theta_{4,6}(\tau,\theta)\over
                              \eta(\tau)}}\,U_i(\tau) &\cr }$$
where the index $i$ runs over all flow invariant orbits with integer $U(1)$
charge and we have expressed the $c=1$ characters in terms of $SU(2)$
$\Theta$-functions
$$\Theta_{n,m}(q,\theta)=\sum_{j\in\BZ+{n\over 2m}}\,
q^{mj^2}\, e^{2\pi i\theta j}.\eqno(2.4) $$
Thus, the series $S_i(\tau), T_i(\tau), U_i(\tau)$ can be interpreted as
generalized string functions corresponding to the coset
$$   {\sigma-{\rm model}\over U(1)}, \eqno(2.5)$$
where the $U(1)$ current $j(z)=\sqrt 3 i \pt\phi(z)$ is compactified on a
circle such that the vertex operator $X^{\pm 3}(z)=\sqrt{6}\,
:e^{\pm i\sqrt 3 \phi(z)}:$ is well-defined. However, by factorizing an
$N=2$ $\sigma$-model as
$$  \sigma{\rm -model}=({\rm compact}\
 U(1))\otimes ({\rm non-supersymmetric\  c=8\  CFT)}\eqno(2.6) $$
we have interchanged in some sense the roles of the two spin-$3/2$ fields
$G^\pm(z)$ and $X^\pm(z)$. In particular, the superpartners $G^\pm(z)$ and
$Y^{\pm 2}(z)$ are now contained in the second and third term of (2.3) with
suitable fields $\Gamma^\pm(z)$ from the non-supersymmetric $c=8$ CFT:
$$  G^\pm(z)=\sqrt 6  :e^{\pm i{1\over \sqrt 3}\phi(z)}:\,
\otimes\, \Gamma^\pm_{H=4/3}(z),\quad\quad
 Y^{\pm 2}(z)= \pm 6 :e^{\pm i{2\over \sqrt 3}\phi(z)}:\,
\otimes\, \Gamma^\pm_{H=4/3}(z) \eqno(2.7)$$
whereas the total $c=9$ energy-momentum tensor remains in the first term
$$ {\ts    L_{\rm tot}={1\over 6}:jj:+L_{c=8} .} \eqno(2.8)$$
These observations thus  reveal a different way of constructing $N=2$
$\sigma$-models. One considers (2.3,2.6,2.7) as the starting point and adds
a suitable rational $c=8$ non-supersymmetric CFT containing a simple current
of dimension $H=4/3$. That the existence of such a current is sufficient for
the entire model to contain the ${\cal{SW}}(1,(3/2)^{\pm 3})$ spectral
flow algebra can be seen by the following argument. Due to the general
formula for a simple current of order $N$ [\sche,\schz]
$$ H={r (N-1)\over 2N }\ {\rm mod}\,\BZ,
      \quad r\in\BZ \eqno(2.9)$$
the order of the simple current $\Gamma^+$ must be divisible by $3$. Thus,
we define $\lb \Gamma^- \rb = \lb \Gamma^+ \rb^{N-1}$. Because of
$  \lb \Gamma^+ \rb\times \lb \Gamma^- \rb = \lb\id\rb$
the conformal dimension of $\Gamma^-$ has to be $H=4/3$, as well. This
implies the OPE to have the following parafermionic like form:
$$  \Gamma^+(z)\,  \Gamma^-(w)={1\over (z-w)^{8\over 3}}+
{1\over 3} {L_{c=8}(w)\over (z-w)^{2\over 3}} +\ldots \eqno(2.10)$$
Note that in general in (2.10) there can also appear spin-1 currents.
However, these can be canceled by choosing the symmetric combination
${1\over 2}( \Gamma^+ +  \Gamma^-)$, which corresponds to modding out the
simple currents $ \Gamma^+$ and $ \Gamma^-$ separately and then adding the
partition functions. We will come back to this point in sect.\ 3. Analogous
to the parafermionic case, (2.10) implies that $G^+(z)\,G^-(w)$ satisfies
the $N=2$ Virasoro relation. The associativity of the OPE and the primarity
of $G^\pm(z)$ then implies that the OPEs  $G^\pm(z)\,G^\pm(w)$ have
vanishing singular parts. The OPE $G^\pm(z)\, X^{\mp 3}(w)$ shows that
$Y^{\pm 2}(z)$ is the superpartner of $X^{\pm 3}(z)$, so that the entire
${\cal{SW}}(1,(3/2)^{\pm 3})$ algebra is satisfied.
\pano
Before continuing the study of these string functions, we briefly review the
construction of heterotic strings, when a $\sigma$-model partition
function has already been chosen [\egu]. Assume, given a $\sigma$-model
partition function in the $NS$ sector:
$$  Z^{NS}=\sum_{ij}\, N_{ij}\, \chi^{\sigma,NS}_i(\tau,\theta)\,
      \o{\chi}^{\sigma,NS}_j(\o{\tau},\o{\theta}) \eqno(2.11)$$
where the characters can be expanded as in (2.3). Then the other sectors can
be calculated easily:
$$\eqalignno{ \chi^{\widetilde {NS}}_i
 (\tau,\theta)&={\ts {\Theta_{0,6}-\Theta_{6,6}\over \eta}}\, S_i(q)
    +{\ts {\Theta_{-4,6}-\Theta_{2,6}\over \eta}}\, T_i(q) +
   {\ts {-\Theta_{-2,6}+\Theta_{4,6}\over \eta} \, U_i(q) }  &\cr
 \chi^{R}_i(\tau,\theta)&={\ts {\Theta_{-3,6}+\Theta_{3,6}
  \over \eta}}\, S_i(q)
    +{\ts {\Theta_{-1,6}+\Theta_{5,6}\over \eta}}\, T_i(q) +
     {\ts {\Theta_{-5,6}+\Theta_{1,6}\over \eta}} \, U_i(q)  &(2.12)\cr
    \chi^{\widetilde R}_i
 (\tau,\theta)&={\ts {-\Theta_{-3,6}+\Theta_{3,6}\over \eta}}\, S_i(q)
    +{\ts {\Theta_{-1,6}-\Theta_{5,6}\over \eta}}\, T_i(q) +
     {\ts {\Theta_{-5,6}-\Theta_{1,6}\over \eta}} \, U_i(q)  .&\cr }$$
The sectors with well-defined $U(1)$ charge parity are
$$\eqalignno{  NS^+_i&=
 {\ts {\Theta_{0,6}\over \eta}\, S_i+{\Theta_{-4,6}\over \eta}\, T_i+
        {\Theta_{4,6}\over \eta} \, U_i \quad\quad (Q\in 2\BZ)} &\cr
  NS^-_i&={\ts {\Theta_{6,6}\over \eta}\, S_i+{\Theta_{2,6}\over \eta}\, T_i+
    {\Theta_{-2,6}\over \eta} \, U_i \quad\quad (Q\in 2\BZ+1)} &(2.13)\cr
  R^+_i&={\ts {\Theta_{3,6}\over \eta}\, S_i+{\Theta_{-1,6}\over \eta}\, T_i+
   {\Theta_{-5,6}\over \eta} \, U_i \hskip-5.5pt
                   \quad\quad (Q\in 2\BZ-1/2)} &\cr
  R^-_i&={\ts {\Theta_{-3,6}\over \eta}\, S_i+{\Theta_{1,6}\over \eta}\, T_i+
   {\Theta_{5,6}\over \eta} \, U_i \quad\quad (Q\in 2\BZ+1/2)}. &\cr }$$
In order to obtain a heterotic string model we have to combine the internal
part with the four-dimensional flat space-time part, their fermionic
partners on the left-moving side and the gauge group $E_8\otimes SO(10)$ on
the right-moving side. How this can be done while preserving modular
invariance has been shown by Gepner [\gepe]. In light cone gauge one first
couples the internal part to the two left-moving fermions which form a $SO(2)$
Kac-Moody algebra admitting the following four integrable representations:
$(\chi_0)^{H=0}_{Q=0}$, $(\chi_v)^{H=1/2}_{Q=1}$, $(\chi_s)^{H=1/8}_{Q=-1/2}$
and $(\chi_c)^{H=1/8}_{Q=1/2}$ where $Q$ is defined modulo $2\BZ$.
Finally, after performing a supersymmetric orbit construction using the
spectral flow operator with $\eta=1/2$ in the internal part, which
corresponds to the Ramond ground state $(H,Q)=(3/8,3/2)$, and the spinor
representation in the external part, one obtains
$$ \chi_{L,i}(\tau,\theta)=\chi^{SO(2)}_v NS^+_i
 + \chi^{SO(2)}_0 NS^-_i - \chi^{SO(2)}_c R^-_i -
             \chi^{SO(2)}_s R^+_i. \eqno(2.14)$$
In the right-moving sector using the bosonic string map, one arrives at
$$ \chi_{R,i}(\tau)= \left( \chi^{SO(10)}_0 NS^+_i +
 \chi^{SO(10)}_v NS^-_i + \chi^{SO(10)}_c R^+_i +
          \chi^{SO(10)}_s R^-_i \right) \chi^{E_8}_0. \eqno(2.15) $$
Using the explicit expressions for the $SO(10)$ characters, the splitting
(2.3) of $\sigma$-model characters and the following expressions for the
$E_6$ characters
$$\eqalignno{ \chi^{E_6}_0&=
 {1\over \eta^6}\left\{ (\theta_{0,6}+\theta_{6,6})\vartheta_3^5 +
              (\theta_{0,6}-\theta_{6,6})\vartheta_4^5 +
            (\theta_{-3,6}+\theta_{3,6})\vartheta_2^5 \right\} &\cr
    \chi^{E_6}_{27}&=
 {1\over \eta^6}\left\{ (\theta_{-5,6}+\theta_{1,6})\vartheta_2^5 +
     (\theta_{-2,6}+\theta_{4,6})\vartheta_3^5 +
       (-\theta_{-2,6}+\theta_{4,6})\vartheta_4^5  \right\} &(2.16)\cr
    \chi^{E_6}_{\o{27}}&=
    {1\over \eta^6}\left\{ (\theta_{-1,6}+\theta_{5,6})\vartheta_2^5 +
           (\theta_{-4,6}+\theta_{2,6})\vartheta_3^5 +
          (\theta_{-4,6}-\theta_{2,6})\vartheta_4^5  \right\} &\cr }$$
one can write the two orbits in the following way:
$$\eqalignno{ \chi_{L,i}(\tau,\theta)&=\chi^{SO(2)}_v
 NS^+_i + \chi^{SO(2)}_0 NS^-_i - \chi^{SO(2)}_c R^-_i -
          \chi^{SO(2)}_s R^+_i &(2.17)\cr
        \chi_{R,i}(\tau)&= \chi^{E_6}_0\, S_i +
  \chi^{E_6}_{27}\, U_i + \chi^{E_6}_{\o{27}}\, T_i . &\cr }$$
Since both the orbit construction and the bosonic map preserve modular
invariance, the partition function of the heterotic string can be written as
$$ Z\sim {1\over | Im(\tau)|\,
|\eta|^4}\sum_{i,j} \, N_{ij}\,  \chi_{L,i}(\tau,\theta)\,
           \o{\chi}_{R,j}(\o{\tau}) \eqno(2.18)$$
where the constant is fixed by the requirement that the vacuum only appears
once in the partition function. Thus, knowing the $\sigma$-model is
sufficient to determine the heterotic string. Especially, the massless
spectrum can be read off directly from (2.17,2.18) and all information is
contained in the string functions $S_i, T_i, U_i$, which appear in orbits of
conformal dimension $H=0$ and $H=1/2$, respectively.
\bigno
\section{3.\ String functions of a Gepner model with 84 generations}
\meno
Although this model has exhaustively been investigated in the past, we want
to illuminate it from our dual point of view again. We will show that this
model contains in a certain sense a broken $E_8$ gauge symmetry of which
only the abelian part $U(1)^8$ appears in the massless spectrum. The
appearance of an $E_8$ symmetry in the massive spectrum was first observed
in [\terry], where only the vacuum character had been investigated. We will
show that the whole partition function can be reconstructed using affine
$E_8$ $\Theta$-functions at level $k=3$.
\pano
For the internal $\ts{c=9}$ theory, one chooses a tensor product of nine
copies of the $\ts{c=1}$ unitary model of the $N=2$ super Virasoro algebra.
In the $NS$ sector there exist three representations labeled by
$\ts{l=m=0,\,\,l=m=1,\,\,l=-m=1}$ with conformal weights $(H,Q)=(0,0)$ and
$(1/6,\pm 1/3)$, respectively. We denote them by $A$,$B$,$C$. Since the
spectral flow cyclically permutes these three, the only flow invariant orbit
containing the vacuum is
$$ \chi^{\sigma}_0=A^9+B^9+C^9 .\eqno(3.1)$$
Some inspection enables one to write this character as
$$ \chi^{\sigma}_0={\ts {\Theta_{0,6}+\Theta_{6,6}\over \eta}\
     {1\over \eta^8}\, \Theta^{E_8}_{0,k=3}(\tau,0) + \ldots} \eqno(3.2)$$
In table 1 we list all other flow invariant orbits and their combinatorial
multiplicities giving the range of the index $i$,
e.g.\ $1680={9\choose3}{6\choose3}$.
\smno
\cl{\vbox{
\hbox{\vbox{\offinterlineskip
\def\tablespace{height2pt&\omit&&\omit&&\omit&\cr}
\def\tablerule{\tablespace\noalign{\hrule}\tablespace}
\hrule\halign{&\vrule#&\strut\hskip0.2cm\hfil#\hfill\hskip0.2cm\cr
\tablespace
& orbit && $(H,Q)$  &&  comb. factor    &\cr
\tablerule\tablerule
& $\chi^i_1=
A^3\,B^3\,C^3$ && $(1,0)$ && $1680$  &\cr
\tablespace
& $\chi^i_2=
A^6\,B^3+B^6\,C^3+C^6\,A^3$ && $({1\over 2},1)$  && $84$  &\cr
\tablespace
& $\chi^i_3=
A^6\,C^3+B^6\,A^3+C^6\,B^3$ && $({1\over 2},-1)$  && $84$  &\cr
\tablespace
& $\chi^i_4=
A^7\,B\,C+B^7\,C\,A+C^7\,A\,B$ && $({1\over 3},0)$  && $72$  &\cr
\tablespace
& $\chi^i_5=
A^4\,B^4\,C+B^4\,C^4\,A+C^4\,A^4\,B$ && $({5\over 6},\pm 1)$  && $630$  &\cr
\tablespace
& $\chi^i_6=
A^5\,B^2\,C^2+B^5\,C^2\,A^2+C^5\,B^2\,C^2$ && $({2\over 3},0)$  && $756$ &\cr
\tablespace}\hrule}}
\hbox{\hskip 0.5cm {\klein table 1: orbits and multiplicities for the
$(1)^9$ model} }}}
\smno
On the level of characters for $j$ fixed all $\chi^i_j$ are identical. Thus,
due to the philosophy of [\egu] after really identifying them one can
calculate the $S$-matrix:
\smno
$$ S={1\over 27\sqrt 3}\left
  (\matrix{ 1 & 1680 & 84  & 84 & 72 & 630 & 756   \cr
                   {1\over 3} & -7 & 1  & 1 & -3 & -6 & 9   \cr
                   1 & 20 & \kappa  & \kappa^*  & 18 & -45 & 27   \cr
                   1 & 20 & \kappa^*  & \kappa  & 18 & -45 & 27   \cr
                   1 & -70 & 21 & 21 & 27 & 0 & 0  \cr
                   1 & -16 & -6  & -6 & 0 & 27 & 0   \cr
                   1 & 20 & 3  & 3 & 0 & 0 & -27   \cr}
\right) \eqno(3.3) $$
\smno
with $\kappa=-21/2+i\,27\sqrt 3/2$. The factor $1/3$ is due to the smaller
orbit length of $\ts{\chi_2^i}$. Thus, the following left-right symmetric
combination is a modular invariant partition function $(5040=3\cdot1680)$:
$$ Z^{NS}\sim | \chi_0 |^2 + 5040\,
| \chi_1 |^2  + 84 \, | \chi_2 |^2 + 84\, | \chi_3 |^2 + 72 \, | \chi_4 |^2
           + 630\, | \chi_5 |^2 + 756\, | \chi_6 |^2  ,\eqno(3.4)$$
leading to $n_{27}=84$ generations and $n_{\o{27}}=0$ antigenerations.
Furthermore, in the massless spectrum there appear $n_1=252$ spin-zero
gauge singlets and eight $U(1)^8$ gauge particles. Apparently from (3.3),
the term $|\chi_2-\chi_3|^2$ is invariant by itself, so that as suggested
in [\egu] multiples of it can be added to (3.4) yielding string models with
a wide range of Euler numbers $\chi/2\in\{ -84, -82,\ldots,82,84 \}$. There
are two consistent arguments which show that only the model (3.4) and its
mirror (i.e.\ $n_{27}=0\,,n_{\o{27}}=84$)
$$   Z^{NS}_M\sim Z^{NS}-84\, |\chi_2-\chi_3|^2, \eqno(3.5)$$
really give consistent string models. Firstly, via a slightly modified
Verlinde formula reflecting $S$ not to be symmetric we have calculated the
fusion rules associated to the $S$-matrix (3.3) which shows that only the
two left-right combinations following from (3.4) and (3.5) satisfy a closed
operator algebra on the tree level. Secondly, the modular invariance of all
these partition functions is an accident resulting from the
information-loosing identification of all the characters in table 1.
Treating them differently, the term added in (3.5) is no longer modular
invariant, but all combinations which are antisymmetric under a $U(1)$ flip
$P:j(z)\to -j(z)$ still form a subrepresentation of the modular group. Thus,
the analogue of (3.5) is
$$  Z^{NS}_M\sim Z^{NS}-\sum_{i,j} | \chi^i_j-\chi^{P(i)}_{P(j)} |^2
\eqno(3.6)$$
which again can be shown to give only the mirror (3.5), for instead of
coupling states with identical $U(1)$ charges, (3.6) couples a state with
charge $Q$ to a state with charge $-Q$. This way of constructing the mirror
of a Gepner-type string model has been suggested in [\gre] and will be
generalized in [\self]. Now, we will show that this model can be rewritten
as $$ \sigma{\rm -model\, of}\,
 (1)^9={\rm compact}\ U(1)\otimes{\rm compact}\
           U(1)^8\ {\rm on}\ 3\,M(E_8) .\eqno(3.7)$$
For the root lattice $M(E_8)$ we choose the following representation:
$$ M(E_8)=\left\{ (x_1,\ldots,x_8)\ {\Bigg{\vert}}
\ \left({\rm all}\  x_i\in\BZ \,\vee\,
 {\rm all}\  x_i\in\BZ+{1\over 2}\right)\wedge
                 \sum x_i\in 2\BZ \right\}. \eqno(3.8)$$
It is known that for a simple Lie algebra $G$ all $\Theta$-functions
$$\Theta^G_{\lambda,k}(\tau,\theta)=\sum_{\gamma\in M+{\lambda\over k} }
q^{k{|\gamma|^2\over 2}}\, e^{2\pi i
             \theta\gamma} \eqno(3.9)$$
form a finite-dimensional representation of the modular group if the
weights are reduced to the finite set $\lambda\in 3M^*/M$ with $M^*$ to be
the dual lattice. In the case of simply-laced Lie algebras the dual lattice
is identical to the weight lattice. In table 2 we list all weights contained
in $3M(E_8)/M(E_8)$ calculating also their conformal dimensions $H$ and
ground state degeneracies.
\smno
\cl{\vbox{
\hbox{\vbox{\offinterlineskip
\def\tablespace{height2pt&\omit&&\omit&&\omit&&\omit&&\omit&\cr}
\def\tablerule{\tablespace\noalign{\hrule}\tablespace}
\hrule\halign{&\vrule#&\strut\hskip0.2cm\hfil#\hfill\hskip0.2cm\cr
\tablespace
& $H$ && $\lambda$  &&  $\#$  && $\#$  total  && deg.     &\cr
\tablerule\tablerule
& $0$ && $(0^8)$ && $1$  &&   $1$  && $1$ &\cr
\tablerule
& ${1\over 3}$ && $(\pm 1,\pm 1,0^6)$ && $112$   &&    && &\cr
&  && $(\pm {1\over 2}^8)\wedge \sum x_i\in 2\BZ$ && $128$ && $240$
&& $1$ &\cr
\tablerule
&  && $(\pm 1^4,0^4)$ && $1120$  &&  &&  &\cr
& ${2\over 3}$ && $(\pm 2,0^7)$ && $16$ &&  && &\cr
& && $(\pm {3\over 2}, \pm {1 \over 2}^7)\wedge \sum x_i\in 2\BZ$
 && $1024$ && $2160$  && $1$  &\cr
\tablerule
&  && $(\pm 1^6,0^2)$ && $1792$   &&  && &\cr
&  && $(+2,+1^2,0^5)$ && $168$   && && &\cr
&  $1$ && $(-2,-1^2,0^5)$ && $168$  && && &\cr
&  && $(+2,-1^2,0^5)/\BZ_3$ && $56$   &&  && &\cr
&  && $(-2,+1^2,0^5)/\BZ_3$ && $56$  && $2240$ && $3$ & \cr
\tablerule
&  && $(\pm 1^8) \wedge  \sum x_i\in 4\BZ-2$ && $128$ && &&   &\cr
&  ${4\over 3}$ && $( +{3\over 2}^3, \pm{1 \over 2}^5)\wedge\sum x_i\in 2\BZ$
&& $896$ &&  && &\cr
&  && $( -{3\over 2}^3, \pm{1 \over 2}^5)\wedge \sum x_i\in 2\BZ$ && $896$ &&
 $1920$ && $9$ &\cr
\tablespace}\hrule}}
\hbox{\hskip 0.5cm {\klein table 2: all $E_8$ weights} }}}
\smno
It turns out that all $\Theta$-functions with the same dimension are
identical as $q$-series, i.e.\ setting $\theta=0$ in (3.9). Identifying
due to [\egu] all these $\Theta$-functions we can calculate the $S$-matrix
for the remaining five characters with dimensions $H\in\{ 0,1/3,2/3,1,4/3\}$:
\smno
$$ S={1\over 81}\left(\matrix{ 1 & 240 & 2160  & 2240 & 1920   \cr
                               1 & 69 &  54  &  -28 &  -96   \cr
                               1 & 6 & -27 &    -28 &   48   \cr
                               1 & -3 & -27  &   53 &  -24   \cr
                               1 & -12 &  54  &  -28 &  -15   \cr }\right)
  ,\eqno(3.10)$$
\smno
where not surprisingly the multiplicities appear in the first row. Note
that due to the identification this $S$-matrix is no longer symmetric,
but $S^2=1$ is still true. Now, we tensor this model with the universal
$c=1$ part and mod out the simple current
$$ Y^{+2}(z)=6 :e^{i{2\over \sqrt 3}\phi(z)}:\,
 \otimes\, :e^{i{1\over \sqrt 3} \beta {\Phi} }:\eqno(3.11) $$
with $\beta=(-1,1,1,1,1,1,1,1)$ and $\Phi$ containing the eight free bosons,
yielding a free field realization of the $c=8$ part. The fusion rules for a
system of $\Theta$-functions are quite simple:
\smno
$$ \lb\Theta_{\mu,k}\rb\times\lb\Theta_{\nu,k}\rb =\lb\Theta_{\mu+\nu,k}\rb
,\eqno(3.12) $$ so that the monodromy charge $Q(\nu)$ of some field $\lb
\id\rb\otimes\lb\Theta_{\nu,k}\rb$ concerning to the simple
current (3.11) is
$$Q(\nu)={|\nu|^2+|\beta|^2-|\nu+\beta|^2\over 6 }\
 {\rm mod}\,\BZ.\eqno(3.13)$$
Forming orbits and projecting them onto those with integer monodromy charge
yields exactly the characters of table 2 including the correct conformal
weights and multiplicities. As announced in the previous section, the eight
spin-one currents also appear in the OPE $G^+(z)\, G^-(w)$, so that we are
led to choose the following symmetrized form of the $\sigma$-model orbits:
$$\eqalignno{\chi_j^i&={\ts {\Theta_{0,6}+\Theta_{6,6}\over \eta}\
   {1\over \eta^8}\, {1\over 2} \left( \Theta^{E_8}_{\lambda^i,k=3}(\tau,0)+
   \Theta^{E_8}_{-\lambda^i,k=3}(\tau,0) \right) } &\cr
     & {\ts +{\Theta_{-4,6}+\Theta_{2,6}\over \eta} \ {1\over \eta^8}\,
  {1\over 2} \left( \Theta^{E_8}_{\lambda^i+\beta,k=3}(\tau,0)+
   \Theta^{E_8}_{-\lambda^i-\beta,k=3}(\tau,0) \right)} &(3.14)\cr
     & {\ts + {\Theta_{-2,6}+\Theta_{4,6}\over \eta} \  {1\over \eta^8}\,
  {1\over 2} \left( \Theta^{E_8}_{\lambda^i+2\beta,k=3}(\tau,0)
     + \Theta^{E_8}_{-\lambda^i-2\beta,k=3}(\tau,0) \right) \quad
         \quad {\rm   mod}\, 3M.} &\cr } $$
For brevity, we present in table 3 only those $E_8$ weights $\lambda$ leading
to the 84 massless orbits
\smno
\cl{\vbox{
\hbox{\vbox{\offinterlineskip
\def\tablespace{height2pt&\omit&&\omit &\cr}
\def\tablerule{\tablespace\noalign{\hrule}\tablespace}
\hrule\halign{&\vrule#&\strut\hskip0.2cm\hfil#\hfill\hskip0.2cm\cr
\tablespace
&  $\lambda^i$  &&  deg.       &\cr
\tablerule\tablerule
& $(0; (-1)^6,0)$ &&  $7$ &\cr
\tablespace
& $(1;(-1)^5,0^2)$  && $21$    &\cr
\tablespace
& $(2;1^2,0^5)$  && $21$    &\cr
\tablespace
& $(0;-2,1^2,0^4)/\BZ_3$  && $35$    &\cr
\tablespace}\hrule}}
\hbox{\hskip 0.5cm {\klein table 3: $E_8$ weights leading to massless orbits
} }}}
\smno
where the first entry should be regarded as fixed in every row. For the 84
antigenerations of the mirror model one has only to exchange plus and minus
signs in table 3. This concludes the reconstruction of the $(1)^9$ Gepner
model in terms of $E_8$ $\Theta$-functions showing that this simple model
already has very intriguing string functions.
\bigno
\section{4.\ A simple toy model}
\meno
We will choose for the non-supersymmetric $c=8$
theory the tensor product of four copies
of the Kac-Moody algebra $SU(3)_1$, which in particular satisfies the
condition of containing a simple current of dimension $H=4/3$.
There exist only three representations of $SU(3)_1$ denoted by
$\chi_0, \chi_3, \chi_{3^*}$ satisfying the following fusion rules:
$$ \lb \chi_3 \rb\times  \lb \chi_0 \rb = \lb \chi_3 \rb, \quad
              \lb \chi_3 \rb\times  \lb \chi_3 \rb= \lb \chi_{3^*} \rb, \quad
              \lb \chi_3 \rb\times  \lb \chi_{3^*} \rb = \lb \chi_0 \rb.
 \eqno(4.1)$$
Thus, $\lb \chi_3 \rb^4$ really is the desired simple current of
$(SU(3)_1)^4$. Modding out the simple current $Y^{+2}(z)$ in
$$\sigma{\rm -model}={\rm compact}\
U(1)\otimes \left(SU(3)_1\right)^4 \eqno(4.2)$$
the orbits listed in table 4 survive.
\smno
\cl{\vbox{
\hbox{\vbox{\offinterlineskip
\def\tablespace{height2pt&\omit&&\omit&&\omit&\cr}
\def\tablerule{\tablespace\noalign{\hrule}\tablespace}
\hrule\halign{&\vrule#&\strut\hskip0.2cm\hfil#\hfill\hskip0.2cm\cr
\tablespace
& orbit $\chi^i_j$ && $(H,Q)$  &&  comb. factor    &\cr
\tablerule\tablerule
& $\chi_0=A\, \chi_0^4+B\, \chi_3^4 +C\, \chi_{3^*}^4$ && $(0,0)$ && $1$ &\cr
\tablespace
& $\chi^i_1=A\, \chi_0\, \chi_3^3+B\,\chi_3\,\chi_{3^*}^3 +C\, \chi_{3^*}\,
 \chi_0^3$ && $({1\over 2},-1)$ && $4$
&\cr
\tablespace
& $\chi^i_2=A\, \chi_0\, \chi_{3^*}^3+B\,\chi_3\,\chi_{0}^3 +C\, \chi_{3^*}\,
 \chi_3^3$ && $({1\over 2},+1)$ && $4$
&\cr
\tablespace
& $\chi^i_3=A\, \chi_0^2\, \chi_3\, \chi_{3^*} + B\,\chi_3^2\,\chi_{3^*}\,
 \chi_0 + C\, \chi_{3^*}^2\, \chi_0\, \chi_3$ && $({2\over 3},0)$ &&
$12$ &\cr
\tablespace
& $\chi^i_4=A\, \chi_3^2\, \chi_{3^*}^2+B\,\chi_{3^*}^2\,\chi_{0}^2 +C\,
\chi_{0}^2\, \chi_3^2$ && $({5\over
6},\pm1)$ && $6$ &\cr
\tablespace}\hrule}}
\hbox{\hskip 0.5cm {\klein table 4: orbits and multiplicities for the
toy model} }}}
\smno
Now $A,B,C$ denote the three characters of the $c=1$ part (2.3), e.g.\ $\ts{
A={\Theta_{0,6}(\tau,\theta)+\Theta_{6,6}(\tau,\theta)\over\eta(\tau)}}$.
Actually, we know that in order to have a good field interpretation one has
to symmetrize the simple current as ${1\over 2}(\lb \chi_3 \rb+\lb \chi_{3^*}
\rb)$. This does not effect, however, the characters in table 4 as
$(\tau,\theta)$ series, so that we immediately obtain the diagonal partition
function, which always guarantees consistency on the tree level, as well:
$$ Z^{NS}\sim | \chi_0 |^2 + 4  | \chi_1 |^2 + 4 | \chi_2 |^2 +
12  | \chi_3 |^2 + 6  | \chi_4 |^2  .\eqno(4.3)$$
Using this $\sigma$-model as the internal part of a heterotic string model
one obtains a vacuum with $n_{27}=36$, $n_{\o{27}}=0$, $n_{1}=324$ and an
obviously enlarged gauge group $E_6\times SU(3)^4$ in the right-moving
sector. The massless fermions $\psi^{ab}_n$ carry three indices, two
of them concerning to this entire gauge group: $a$ for the fundamental
$27$-dimensional representation of $E_6$ and $b$ for the fundamental
three-dimensional representation of in each case one of the four $SU(3)$
factors. With respect to the remaining three $SU(3){\rm s}$, the fermions are
singlets. Due to the degeneracy of $\chi_3(q)=3q^{1\over3}+\ldots$ the third
index $n\in\lbrace 1,2,3\rbrace$ just counts the particles in the
left-moving sector. We have found a model with this massless spectrum and
extended gauge group in [\lerz] in the context of the covariant lattice
approach, from which it is absolutely not obvious that it can be written
in such a simple way. In addition, the number of (anti)generations occurs
as an orbifold of the $(1)^9$ Gepner model [\kreuz].
\bigno
\section{5.\ Conclusion and outlook}
\meno
In this letter we have outlined an alternative way of constructing $N=2$
Calabi-Yau $\sigma$-model partition functions. We have both reconstructed
one already known model realizing a broken $E_8$ symmetry and used the dual
approach as a constructive one. In principle every $c=8$ non-supersymmetric
CFT containing a simple current of dimension $H=4/3$ can be used as string
functions in this formalism. Whether or not new three generation models can
be achieved remains to be seen. Furthermore, it should be possible to
transform other methods like the orbifold construction into  our approach.
Since the $c=1$ part is universal for all $\sigma$-models, one might hope
this construction scheme to be more suitable for the investigation of
marginal deformations of the CFT, reflecting deformations of the complex and
K\"ahler structure of the underlying Calabi-Yau manifold.
\smno
{\bf Acknowledgements}
\pano
It is a pleasure to thank L.\ Dolan,  W.\ Nahm, R.\ Schimmrigk and
M.\ Terhoeven for discussion. This work is supported by U.S.\ DOE grant No.\
DE-FG05-85ER-40219.
\bigno
\section{References}
\meno
\baselineskip=10.7pt
\bibitem{\abel} S.A.\ Abel, C.M.A.\ Scheich,
{\it Classification of $(2,2)$ compactifications by free fermions $2$ },
Phys.\ Lett.\ {\bf B333} (1994) 68
\bibitem{\ant} I.\ Antoniadis, C.\ Bachas, C.\ Kounnas,
{\it Four-dimensional superstrings},
Nucl.\ Phys.\ {\bf B289} (1987) 87
\bibitem{\banks} ${\rm T.\,Banks, L.J.\,Dixon, D.Friedan, E.\,Martinec}$,
{\it Phenomenology and conformal field theory, or Can string theory predict
the weak mixing angle{\rm ?}},
${\rm Nucl.\,Phys.\,{\bf B299}}\,(1988)\,613$
\bibitem{\self} R.\ Blumenhagen, A.\ Wi{\ss}kirchen, {\it in preparation}
\bibitem{\cande} P.\ Candelas, G.T.\ Horowitz, A.\ Strominger, E.\ Witten,
{\it Vacuum configurations for superstrings},
Nucl.\ Phys.\ {\bf B258} (1985) 46
\bibitem{\candz} P.\ Candelas, M.\ Lynker, R.\ Schimmrigk,
{\it Calabi-Yau manifolds in weighted $P(4)$},
Nucl.\ Phys.\ {\bf B341} (1990) 383
\bibitem{\candd} P.\ Candelas, E.\ Derrick, L.\ Parkes,
{\it Generalized Calabi-Yau manifolds and the mirror of a rigid manifold},
Nucl.\ Phys.\ {\bf B407} (1993) 115
\bibitem{\egu} T.\ Eguchi, H.\ Ooguri, A.\ Taormina, S.K.\ Yang,
{\it Superconformal algebras and string compactification on manifolds
with $SU(n)$ holonomy},
Nucl.\ Phys.\ {\bf B315} (1989) 193
\bibitem{\fuchs} J.\ Fuchs, A.\ Klemm, C.M.A.\ Scheich, M.G.\ Schmidt,
{\it Spectra and symmetries of Gepner models compared to Calabi-Yau
compactifications},
Ann.\ Phys.\ {\bf 204} (1990) 1
\bibitem{\gepe} D.\ Gepner, {\it Space-time supersymmetry in
compactified string theory and superconformal models},
Nucl.\ Phys.\ {\bf B296} (1988) 757
\bibitem{\gepz} D.\ Gepner, {\it String theory on Calabi-Yau manifolds: the
three generation case},
preprint PUPT-0085, December 1987
\bibitem{\gre} B.R.\ Greene, M.R.\ Plesser, {\it Duality in Calabi-Yau
moduli space}, Nucl.\ Phys.\ {\bf B338} (1990) 15
\bibitem{\gros} D.J.\ Gross, J.A.\ Harvey, E.\ Martinec, R.\ Rohm,
{\it Heterotic string theory $(I)+(II)$},
Nucl.\ Phys.\ {\bf B256} (1985) 253, Nucl.\ Phys.\ {\bf B267} (1985) 75
\bibitem{\kaw} H.\ Kawai, D.C.\ Lewellyn, S.-H.$\,$H.\ Tye,
{\it  Construction of fermionic string models in four dimensions},
Nucl.\ Phys.\ {\bf B288} (1987) 1
\bibitem{\kreuz} M.\ Kreuzer, H.\ Skarke,
{\it $ADE$ string vacua with discrete torsion},
Phys.\ Lett.\ {\bf B318} (1993) 305
\bibitem{\lere} W.\ Lerche, D.\ L\"ust, A.N.\ Schellekens,
{\it Chiral four-dimensional heterotic strings from self-dual lattices},
Nucl.\ Phys.\ {\bf B287} (1987) 477
\bibitem{\lerz} ${\rm W.\,Lerche, A.N.\,Schellekens, N.\,Warner}$,
{\it Lattices and strings},
${\rm Phys.\,Rep.\,{\bf 177}}\,(1989)\,1$
\bibitem{\oda} S.\ Odake, {\it Character formulas of an
extended superconformal algebra relevant to string compactification},
Int.\ J.\ Mod.\ Phys.\ {\bf A5} (1990) 897
\bibitem{\sche} A.N.\ Schellekens, S.\ Yankielowicz,
{\it Extended chiral algebras and modular invariant partition functions},
Nucl.\ Phys.\ {\bf B327} (1989) 673
\bibitem{\schz} A.N.\ Schellekens, S.\ Yankielowicz,
{\it Modular invariants from simple currents. An explicit proof},
Phys.\ Lett.\ {\bf B227} (1989) 387
\bibitem{\schd} A.N.\ Schellekens, S.\ Yankielowicz,
{\it New modular invariants for $N=2$ tensor products and four-dimensional
strings}, Nucl.\ Phys.\ {\bf B330} (1990) 103
\bibitem{\rolfe} R.\ Schimmrigk,
{\it A new construction of a three-generation Calabi-Yau
manifold},
Phys.\ Lett.\ {\bf B193} (1987) 175
\bibitem{\rolfz} R.\ Schimmrigk,
{\it Critical superstring vacua from noncritical manifolds: a novel
framework for string compactification},
Phys.\ Rev.\ Lett.\ {\bf 70} (1993) 3688
\bibitem{\rolfd} R.\ Schimmrigk,
{\it Mirror symmetry and string vacua from a special class of Fano
varieties}, preprint BONN-TH-94-07, May 1994, hep-th/9405086
\bibitem{\terry}  M.\ Terhoeven,
{\it Gepner models and $\cal W$-algebras},
    diploma thesis, BONN-IR-92-20, June 1992  (in german)
\vfill
\end